\documentclass[
 amsmath,amssymb,
 aps,
pra,
twocolumn
]{revtex4-1}

\usepackage{graphicx}
\usepackage{dcolumn}
\usepackage{bm}
\usepackage[unicode=true, breaklinks=false, pdfborder={0 0 1}, backref=false, colorlinks=true, linkcolor=blue, urlcolor=blue, citecolor=blue]{hyperref}
\usepackage{braket}
\usepackage{bbm}

\usepackage{txfonts}
\usepackage[usenames, dvipsnames]{color}
\usepackage{soul}
\usepackage[normalem]{ulem}

\allowdisplaybreaks

\newcommand{\tr}{\operatorname{tr}}
\newcommand{\id}{\mathbbm{1}}

\begin{document}

\title{Objectivity (or lack there of): a comparison between predictions of quantum Darwinism and spectrum broadcast structure}

\author{Thao P. Le}

\email{thao.le.16@ucl.ac.uk}

\affiliation{Dept. of Physics and Astronomy, University College London, Gower Street, London WC1E 6BT}

\author{Alexandra Olaya-Castro}

\affiliation{Dept. of Physics and Astronomy, University College London, Gower Street, London WC1E 6BT}

\begin{abstract}
Quantum Darwinism and spectrum broadcast structure describe the emergence of objectivity in quantum systems. However, it is unclear whether these two frameworks lead to consistent predictions on the objectivity of the state of a quantum system in a given scenario. In this paper, we jointly investigate quantum Darwinism and spectrum broadcasting, as well as the subdivision of quantum Darwinism into accessible information and quantum discord, in a two-level system interacting with an $N$-level environment via a random matrix coupling. We propose a novel partial trace method to suitably and consistently partition the effective $N$-level environment, and compare the predictions with those obtained using the partitioning method proposed by Perez [Phys. Rev. A 81, 052326 (2010)]. We find that Quantum Darwinism can apparently emerge under the Perez trace even when spectrum broadcast structure does not emerge, and the majority of the quantum mutual information between system and environment fractions is in fact quantum in nature. This work therefore shows there can be discrepancies between quantum Darwinism, and the nature of information and spectrum broadcast structure.
\end{abstract}

\pacs{Valid PACS appear here}

\maketitle

\section{Introduction}

In the most general scenario, a quantum system interacts with an environment with a large number of degrees of freedom. This interaction influences the dynamical evolution of the system, the measurement outcomes of the system observables, and the extent to which quantum properties, such as quantum superpositions and interferences, are suppressed \cite{Zurek2003,Schlosshauer2005}. Through such system-environment interactions, the state of the system can then appear classically objective to different observers \cite{Zurek2003}. How this so-called quantum to classical transition emerges is not fully understood.

Decoherence theory has provided an important framework to explain how quantum superposition states are destroyed and the time scales in which such quantum information is lost to the environment \cite{Zurek2003,Zurek1993,Schlosshauer2007,Joos2003}. Here the role of the environment is restricted to how it affects the system. However, environments are generally non-monolithic: made up of individual photons, spins, or quasi-particles. Observers are usually able to access a \emph{part} of the environment, and hence possibly learn some information about the system. From this perspective the environment of quantum system can be understood as a ``communication channel" between the system and an observer. These ideas have led to the concept of \emph{quantum Darwinism}: during the decoherence process, information about the system is duplicated into different parts environment \cite{Zurek2009, Ollivier2004,Blume-Kohout2006,Blume-Kohout2005,Blume-Kohout2008}. A state is considered \emph{objective} (or \emph{inter-subjective} according to Ref.~\cite{Mironowicz2017}) when different observers can independently access and measure different parts of the environment and independently obtain (the same) information about the system \cite{Ollivier2004,Zurek2009,Horodecki2015}. Quantum Darwinism assesses this by examining the mutual information between system state and environment fragments. It has been shown that the core process of quantum Darwinism is universal to all quantum dynamics, assuming that the environment is sufficiently large \cite{Brandao2015}.

The duplication of information from system to environment can also be viewed as a form of information broadcasting whereby state objectivity is evaluated by looking at the specific system-environment state structure \cite{Horodecki2015}: a state is objective if the system and some parts of the environment have a \emph{spectrum broadcast structure} corresponding to a statistical mixture of distinguishable separable states of the system and each environment fraction. This is a different, and more stringent, definition for objectivity than quantum Darwinism. Given a spectrum broadcast structure for the system-environment state, the ability of different observers to independently determine the properties of the system immediately follows.

Both quantum Darwinism and spectrum broadcasting have been explored, albeit separately, in a number of different specific models  \cite{Tuziemski2015,Tuziemski2015a,Galve2016,Paz2009,Blume-Kohout2008,Balaneskovic2015,Balaneskovic2016,Blume-Kohout2005,Giorgi2015,Lampo2017,Mironowicz2017,Korbicz2014,Pleasance2017,Riedel2010,Zwolak2009}. However, we identify two key issues: First, it is not clear which dynamics hinders or aids the emergence of quantum Darwinism. For example, from the non-Markovian perspective, it has been found that strong system-environment interactions, memory effects, and initial correlations can hinder objectivity in particular scenarios \cite{Galve2016,Giorgi2015,Balaneskovic2015,Pleasance2017}, yet not in others \cite{Lampo2017}. Second, there has been no rigorous study of the consistency (or deviation) between predictions on objectivity from quantum Darwinism and spectrum broadcasting. Furthermore, quantum Darwinism and spectrum broadcast structure have mostly been investigated in quantum scenarios where environments have discrete, explicit subsystems. Yet, if quantum Darwinism (and perhaps spectrum broadcasting) is indeed universal to all quantum dynamics, or at least to decohering dynamics, then objectivity should also emerge with a single environment \cite{Perez2010}. For instance, in recent approaches to simulate open quantum systems with photonic qubits \cite{Salvail2013,Liu2011}, the polarization degree of freedom is taken as the system of interest. The frequency or spatial degrees of freedom then acts as the environment---these are continuous degrees of freedom and are not obviously discrete. However, by suitable discretisation, we could define subenvironments and thus apply quantum Darwinism.

In this paper, we show disagreement between the conclusions that one can draw about state objectivity using quantum Darwinism and spectrum broadcast structure in the regime where a quantum dynamics deviates from being Markovian.  We illustrate this with a random matrix model of a two-level system interacting with a $N$-level environment. To identify effective fragments in this environment we use the partial trace method given in Ref.~\cite{Perez2010}. We also propose a different partial trace method that avoids some caveats we identified in the former method. Using the two different trace methods, we investigate quantum Darwinism and spectrum broadcast structure. We find that the conclusions drawn by quantum Darwinism can be inconsistent with those of the accessible information and quantum discord, and in turn with spectrum broadcast structure under the partial trace method of Ref.~\cite{Perez2010}; otherwise, we find that quantum Darwinism is non-applicable under the type of environment implied by this partial trace.

\section{System-environment model \label{sec:Models}}

Random matrix models and random matrix theory have been used to model spectral fluctuations for decades \cite{Brody1981}. Random matrix models have also been used to explore decoherence \cite{Gorin2008,Carrera2014,Lebowitz2015} and quantum chaos \cite{Weidenmueller2009}. If the interaction between system and environment varies rapidly, or if the environment is highly complex, then the coupling can be approximated with a random Gaussian matrix \cite{Bulgac1998,Lutz1999}. This motivates us to consider a model comprising of a two-level spin system interacting with an $N$-level environment via a random Gaussian matrix coupling.

The system Hamiltonian and environment Hamiltonians are, respectively ($\hbar=1$):
\begin{equation}
\hat{H}_{\mathcal{S}}=\dfrac{\Delta E}{2}\sigma_{z},\qquad\hat{H}_{\mathcal{E}}=\sum_{n=0}^{N-1}\varepsilon_{n}\ket{n}\bra{n},
\end{equation}
where the environment is an $N$-level system consisting of levels spaced $\delta\varepsilon/\left(N-1\right)$ apart, ranging consecutively from $\varepsilon_0 = -\delta\varepsilon/2$ to $\varepsilon_{N-1} = \delta\varepsilon/2$.  We consider the following interaction Hamiltonian between system and environment:
\begin{align}
\hat{H}_{SE} & =\sigma_{x}\otimes\lambda R,
\end{align}
where $R=X/\sqrt{8N}$, where $X$ is a Gaussian orthogonal random matrix of size $N$ (a real symmetric matrix with $X_{ij}\sim \mathcal{N}\left(0,1\right)$ and $X_{ii}\sim\sqrt{2}\mathcal{N}\left(0,1\right)$ where $\mathcal{N}\left(0,1\right)$ refers to the normal distribution).
Note that the $\sqrt{8N}$ factor is introduced so that the width of the averaged smooth density of states of $R$ is fixed to unity \cite{Esposito2003}. The choice of using a Gaussian orthogonal ensemble (GOE) coupling rather than a Gaussian unitary ensemble (GUE) is to provide a smoother transition between this work and preceding works by other authors, and to be able to use their results into this study \cite{Perez2010,Esposito2010}.

By increasing the strength of the interaction $\lambda$ (relative to the system and environment energy scales $\Delta E$ and $\delta\varepsilon$ respectively) and by decreasing the number of levels $N$ in the environment, the system-environment correlations strengthen and entropy production increasingly deviates from Markovian predictions \cite{Esposito2010}.

\begin{figure*}
\begin{centering}
\includegraphics[width=1\textwidth]{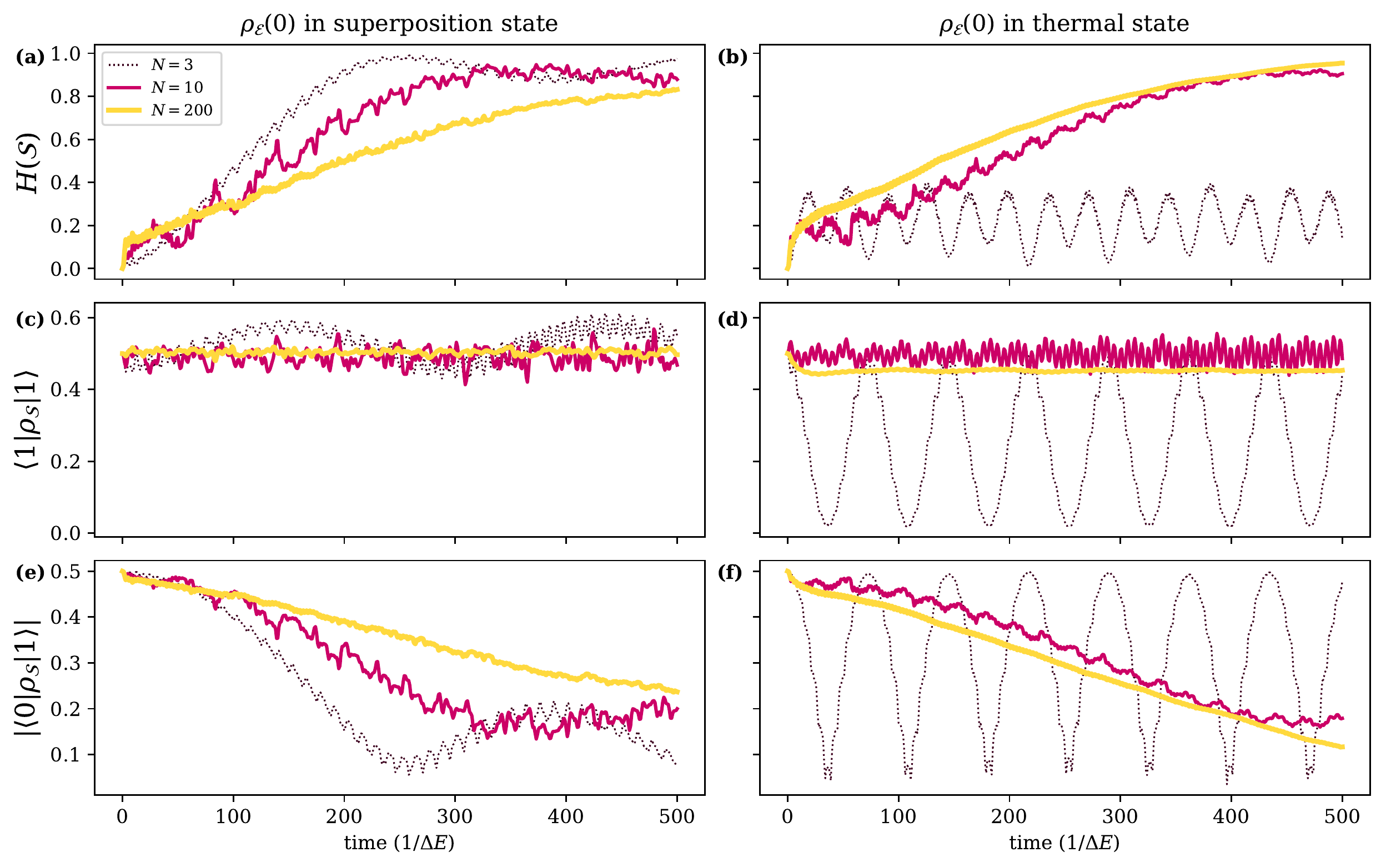}
\par\end{centering}
\caption{\textbf{Evolution of the system as the environment changes}. For all figures the parameters are $\Delta E=1$, $\delta\varepsilon=\Delta E$, $\lambda=\Delta E/5$, and $\beta=10$ ($\hbar=1$). The environment begins in either the superposition state, or in the thermal state with inverse temperature $\beta$. The system begins in the superposition state.
\textbf{(a), (b)} Plots of the system entropy $H\left(\mathcal{S}\right)$,  (\textbf{c), (d)} the excited coefficient of the system, $\braket{1|\rho_{\mathcal{S}}|1}$, and \textbf{(e), (f)}  the absolute value of the off-diagonal coefficient $\left|\braket{0|\rho_{\mathcal{S}}|1}\right|$. As $N$ increases, the interaction between system and environment weakens and the dynamics become increasingly Markovian. \label{fig:evolution}}
\end{figure*}

To illustrate this, we fix the parameters to be $\Delta E=1$, $\delta\varepsilon=\Delta E$, $\lambda=\Delta E/5$, and change $N$, as shown in Fig. \ref{fig:evolution}. The initial state of the system-environment is separable, $\rho_{\mathcal{SE}}\left(0\right)~=~\rho_{\mathcal{S}}\left(0\right)\otimes\rho_{\mathcal{E}}\left(0\right)$. 
The system is initially in superposition state $\rho_{\mathcal{S}}\left(0\right)=\ket{\Psi_{\mathcal{S}}\left(0\right)}\bra{\Psi_{\mathcal{S}}\left(0\right)}$, where $\ket{\Psi_{\mathcal{S}}\left(0\right)}=\frac{1}{\sqrt{2}}\left(\ket{0}+\ket{1}\right)$.
The environment $\rho_{\mathcal{E}}\left(0\right)$ is either in a quantum superposition state, $\rho_{\mathcal{E}}\left(0\right)=\ket{\Psi_{\mathcal{E}}\left(0\right)}\bra{\Psi_{\mathcal{E}}\left(0\right)}$ with $\ket{\Psi_{\mathcal{E}}\left(0\right)}=\frac{1}{\sqrt{N}}\sum_{n=0}^{N-1}\ket{n}$,
or in a thermal state with inverse thermal energy scale $\beta$ so that $\rho_{\mathcal{E}}\left (0 \right )=e^{-\beta\hat{H}_{\mathcal{E}}}/ \tr  [ e^{-\beta\hat{H}_{\mathcal{E}}}]$. We set $\beta~=~10$.
The comparison of the quantum dynamics for different initial states of the environment is of importance as it has been shown that quantum Darwinism is hindered when the environment is in a statistical mixture \cite{Giorgi2015,Zwolak2009,Zwolak2010,Balaneskovic2015,Balaneskovic2016}.

Quantum Darwinism  demands the analysis of the quantum mutual information between system and environment, hence the quantum full dynamics of $\rho_{\mathcal{SE}}\left(t\right)$ is required (aside from specific cases such as with Gaussian states \cite{Blume-Kohout2008}). The joint system-environment at time $t$ is obtained by numerically evolving the initial state via Schrodinger's equation. To do so we use the  QuTip package \cite{Johansson2012,Johansson2013}. The reduced system state is then recovered via the usual trace,
\begin{equation}
\rho_{\mathcal{S}}\left(t\right)=\tr_{\mathcal{E}}\left[\rho_{\mathcal{SE}}\left(t\right)\right].
\end{equation}

The dynamics of the system is displayed in Fig. \ref{fig:evolution}.
For various environment sizes $N=3,10,200$, we plot the system's von Neumann entropy $H\left(\mathcal{S}\right)=H\left(\rho_{\mathcal{S}}\left(t\right)\right)=-\tr\rho_{\mathcal{S}}\left(t\right)\log_{2}\rho_{\mathcal{S}}\left(t\right)$, the excited population $\braket{1|\rho_{\mathcal{S}}\left(t\right)|1}$, and the absolute value of the off-diagonal component $\left|\braket{0|\rho_{\mathcal{S}}\left(t\right)|1}\right|$ corresponding to the coherence of the system state. As $N$ increases, the entropy of system shows roughly monotonic rise and decoherence (relative to the system Hamiltonian) is roughly monotonic (up to small-scale oscillations). Whilst for small $N$ the system shows cycles of gaining and losing entropy and coherence, thereby departing from Markovian dynamics. A thermal environment causes faster decay of the system excited population. The thermal environment also damps out  the small-scale oscillations in entropy, excited state population and decoherence dynamics.

In the model we consider, the system-environment coupling does not commute with the system Hamiltonian. Furthermore, the two energy scales are comparable and therefore the system Hamiltonian cannot be neglected. This implies that the pointer states that are selected by the environment during the decoherence process are not eigenstates of $\sigma_x$ but rather the set of pure system states that are less prone to evolve into an statistical mixture \cite{Zurek1993a}. Later in the paper in section \ref{sec:Spectrum-Broadcast-Structure} when the pointer basis is needed, we show that it can be defined as the system basis in which the shared system-environment information is maximised.

As shown in Ref.~\cite{Esposito2010} the dynamics approaches the Markovian limit as $N$ is increased. Hence, at finite $N$, there are non-negligible system-environment correlations. Therefore, for the remainder of the paper, we fix $N=10$, in order to account for non-Markovian dynamics with a reasonable number of levels in the environment that will allow us to study Quantum Darwinism and spectrum broadcasting.

\section{Quantum Darwinism \label{sec:Quantum-Darwinism}}

Quantum Darwinism is typically studied by considering the mutual information $I\left(\mathcal{S}:\mathcal{F}\right)=H\left(\mathcal{S}\right)+H\left(\mathcal{F}\right)-H\left(\mathcal{SF}\right)$ between the system $\mathcal{S}$ and varying sized fragments of the environment $\mathcal{F}\subseteq\mathcal{E}$. When the mutual information takes value $I\left(\mathcal{S}:\mathcal{F}_{\delta}\right)=(1-\delta) H(\mathcal{S})$, the fragment $\mathcal{F}_{\delta}$ is said to contain roughly all the information of the system state. If this occurs at sufficiently small fractions $\left|\mathcal{F}_{\delta}\right|= f_{\delta}\left|\mathcal{E}\right|$, then we say that there are multiple copies of the information in the environment, and so the system state is objective. In a plot of $I\left(\mathcal{S}:\mathcal{F}\right)$ versus fraction size $f$, this emerges as a ``mutual information plateau'' \cite{Zurek2003}.

In order to define the fragments $\mathcal{F}$, we use two different partial trace methods. The first method the level-partitioning and elimination or P\'erez method \cite{Perez2010}. We also introduce a novel method that does not have the caveats of the first method (explained later in this section). We call our method the staircase environment trace.

The two partitioning methods give different predictions as to whether quantum Darwinism has emerged or not. To examine the discrepancy further, we investigate the mutual information's contributing terms: the accessible ``classical'' information and the quantum discord. Both methods agree that the majority of the mutual information between system and environment is comprised on quantum discord, which alternatively suggests that objectivity has \emph{not} arisen.

\subsection{Partitioning the environment via alternative partial trace methods}

\begin{figure}
\begin{centering}
\includegraphics[width=0.9\columnwidth]{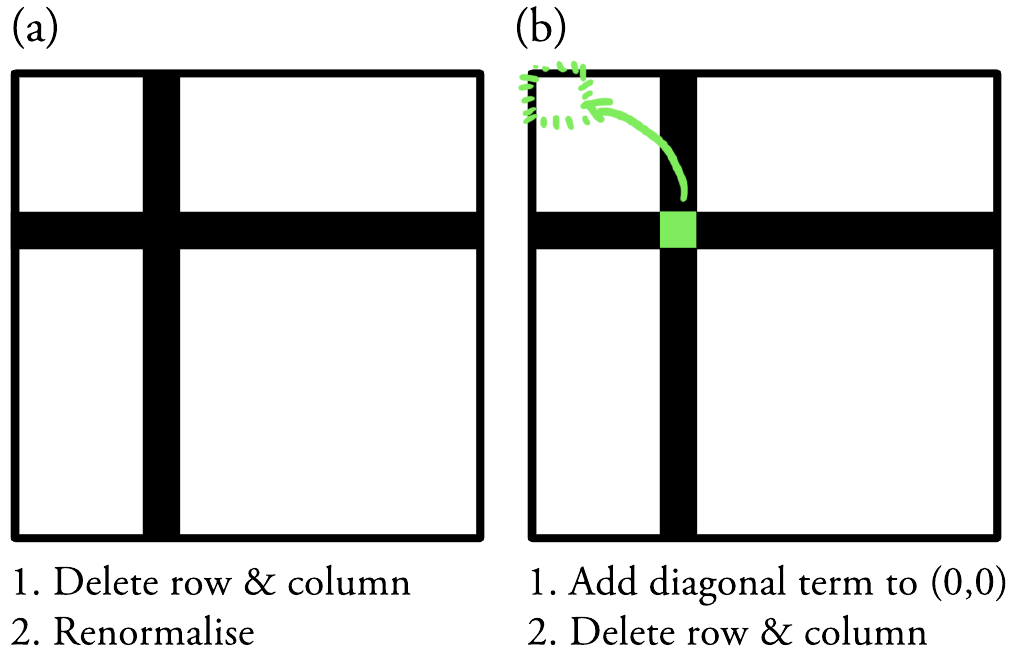}
\par\end{centering}
\caption{\textbf{Graphical depiction of two alternative partial traces}: the modification of a matrix block. \textbf{(a)} The level-partitioning and elimination partial trace method given by \citet{Perez2010}. \textbf{(b)} The staircase environment trace we introduce, based on the assumption that the environment comprises of $N-1$ different subsystems with increasing excited energy. \label{fig:partial_traces}}
\end{figure}

First, we will re-introduce the level-partitioning and elimination partial trace method defined by \citet{Perez2010}, followed by our novel ``staircase environment'' partial trace. Graphical depictions of both partial traces are shown in Fig. \ref{fig:partial_traces} that give a picture of how the associated matrix structure changes due to these partial traces.

\subsubsection{The P\'{e}rez trace (level-partitioning and elimination)}\label{subsubsec:PerezTrace}

The \citet{Perez2010} partial trace method assumes that if we can only access some fraction of levels $\mathcal{F}$, all other levels in the environment are essentially non-existent. Therefore, for a general system-environment state (with implicit time-dependence on the coefficients),
\begin{align}
\rho_{\mathcal{SE}}\left(t\right) & =\sum_{i,j=0,1}\sum_{n,m=0}^{N-1}c_{ijnm}\ket{i}_{\mathcal{S}}\bra{j}\otimes\ket{n}_{\mathcal{E}}\bra{m},\label{eq:general_SE_state}
\end{align}
the P\'{e}rez trace gives the following reduced system-fragment state:
\begin{align}
\rho_{\mathcal{SF}}\left(t\right) & =\dfrac{1}{N_{\mathcal{F}}}\sum_{i,j=0,1}\sum_{n,m\in\mathcal{F}}c_{ijnm}\ket{i}_{\mathcal{S}}\bra{j}\otimes\ket{n}_{\mathcal{E}}\bra{m},\label{eq:PerezTrace}
\end{align}
where $N_{\mathcal{F}}=\sum_{i=0,1}\sum_{n\in\mathcal{F}}c_{iinn}$ is a normalisation factor. Here, $\mathcal{F}\subseteq\left\{ 0,1\ldots,N-1\right\} =\mathcal{E}$ is a subset of possible energy levels. If $N_{\mathcal{F}}=0$, then $\rho_{\mathcal{SF}}\left(t\right)=0$, which corresponds to the environment being in one of the $\mathcal{E}\backslash\mathcal{F}$ levels.

With this method the full system-environment state is pure $\rho_{\mathcal{SE}}\left(t\right)=\ket{\Psi_{\mathcal{SE}}\left(t\right)}\bra{\Psi_{\mathcal{SE}}\left(t\right)}$,
(e.g. $\ket{\Psi_{\mathcal{SE}}\left(t\right)}=\sum_{i=0,1}\sum_{n=0}^{N-1}a_{in}\left(t\right)\ket{i}_{\mathcal{S}}\otimes\ket{n}_{\mathcal{E}}$)
then all system-fragment states are \emph{also} pure $\rho_{\mathcal{SF}}\left(t\right)=\ket{\Psi_{\mathcal{SF}}\left(t\right)}\bra{\Psi_{\mathcal{SF}}\left(t\right)}$,
with
\begin{align}
\ket{\Psi_{\mathcal{SF}}\left(t\right)} &=\dfrac{1}{\sqrt{N_{\mathcal{F}}\left(t\right)}}\sum_{i=0,1}\sum_{n\in\mathcal{F}}a_{in}\left(t\right)\ket{i}_{\mathcal{S}}\otimes\ket{n}_{\mathcal{E}}.
\end{align}

One obvious caveat is that the apparent reduced system state is not the true system state in general. Partial traces should satisfy:
\begin{align}
\rho_{\mathcal{S}} & =\tr_{\mathcal{E}}\left[\rho_{\mathcal{SE}}\right]=\tr_{\mathcal{F}}\left[\rho_{\mathcal{SF}}\right].\label{eq:partial_trace_sys}
\end{align}
However, in general, $\tr_{\mathcal{F}}\left[\rho_{\mathcal{SF}}\right]\neq\rho_{\mathcal{S}}=\tr_{\mathcal{E}}\left[\rho_{\mathcal{SE}}\right]$ if using the P\'{e}rez trace. As such, we also suggest a different partial trace method that does not have this problem.

\subsubsection{The staircase environment trace}

The staircase environment trace assumes that the environment is comprised of $N-1$ two-level subsystems, where each subsystem has increasing energy:
\begin{align}
H_{\mathcal{E}} & =\varepsilon_{0}\ket{0}\bra{0}^{\otimes N-1}+\sum_{n=1}^{N-1}\varepsilon_{n}\ket{1}_{\mathcal{E}_{n}}\bra{1}\otimes\id_{\mathcal{E}\backslash\mathcal{E}_{n}}.
\end{align}
For example, $\ket{1}=\ket{1}_{\mathcal{E}_{1}}\otimes\ket{0}_{\mathcal{E}_{2}}\otimes\cdots\otimes\ket{0}_{\mathcal{E}_{N-1}}$.
Under this assumption, the effective $N$-levels that we are working with are a \emph{subspace} of a larger $2^{N-1}$ environment Hilbert space. For a general system-environment state given in Eq. (\ref{eq:general_SE_state}), tracing out some environment $\mathcal{E}_{k}$ corresponds to 
\begin{align}
\rho_{S\left(\mathcal{E}\backslash\mathcal{E}_{k}\right)}\left(t\right) & =\tr_{\mathcal{E}_{k}}\left[\rho_{\mathcal{SE}}\left(t\right)\right]\nonumber \\
 & =\braket{0_{\mathcal{E}_{k}}|\rho_{\mathcal{SE}}\left(t\right)|0_{\mathcal{E}_{k}}}+\braket{1_{\mathcal{E}_{k}}|\rho_{\mathcal{SE}}\left(t\right)|1_{\mathcal{E}_{k}}}\nonumber \\
 & =\sum_{i,j=0,1}\sum_{\stackrel{n,m=0}{n,m\neq k}}^{N-1}c_{ijnm}\ket{i}_{\mathcal{S}}\bra{j}\otimes\ket{n}_{\mathcal{E}\backslash\mathcal{E}_{k}}\bra{m}\nonumber \\
 & \phantom{=}+\sum_{i,j=0,1}c_{ijkk}\left(t\right)\ket{i}_{\mathcal{S}}\bra{j}\otimes\ket{0}_{\mathcal{E}\backslash\mathcal{E}_{k}}\bra{0}.
\end{align}
A reduced system-fragment state is therefore
\begin{align}
\rho_{\mathcal{SF}}\left(t\right) & =\sum_{i,j=0,1}\sum_{n,m\in\mathcal{F}\cup\left\{ 0\right\} }c_{ijnm}\left(t\right)\ket{i}_{\mathcal{S}}\bra{j}\otimes\ket{n}_{\mathcal{E}\backslash\mathcal{F}}\bra{m}\nonumber \\
 & \phantom{=}+\sum_{i,j=0,1}\sum_{k\in\mathcal{E}\backslash\mathcal{F}}c_{ijkk}\left(t\right)\ket{i}_{\mathcal{S}}\bra{j}\otimes\ket{0}_{\mathcal{E}\backslash\mathcal{F}}\bra{0}.\label{eq:staircasetrace}
\end{align}

The advantage of this method is that the reduced system state as derived from $\rho_{\mathcal{SF}}\left(t\right)$ is equivalent to the true system state derived from $\rho_{\mathcal{SE}}\left(t\right)$, \emph{i.e.}, Eq. (\ref{eq:partial_trace_sys}) holds. The caveat is that we have made the assumption of the excited energy levels for the environment sub-components.

Note that the von Neumann entropy of the reduced state $\rho_{\mathcal{F}}$ or $\rho_{\mathcal{SF}}$ in the $N$-dimensional subspace has the same non-zero eigenvalues as the full representation.

\subsection{Results\label{subsec:qD_results}}

\begin{figure*}
\begin{centering}
\includegraphics[width=1\textwidth]{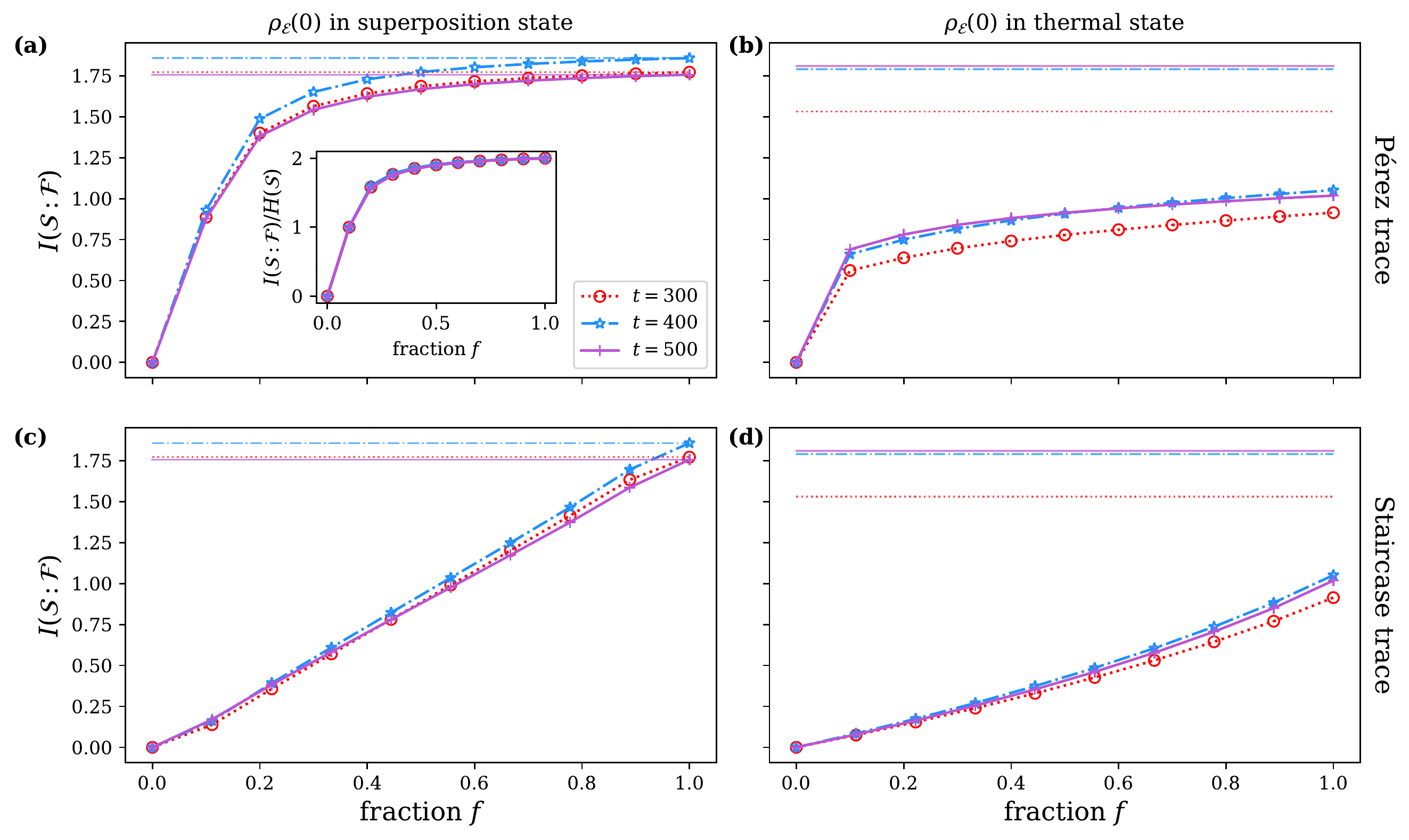}
\par\end{centering}
\caption{\textbf{Mutual information $I\left(\mathcal{S}:\mathcal{F}\right)$ between system and environment fragments.} For all figures the parameters are $\Delta E=1$, $\delta\varepsilon=\Delta E$, $\lambda=\Delta E/5$, and $\beta=10$ for the thermal state ($\hbar=1$). \textbf{(a) and (b)} The \citet{Perez2010} trace (Eq. (\ref{eq:PerezTrace})) is used to form different fractions of the environment. \textbf{(c) and (d)} The staircase trace (Eq. (\ref{eq:staircasetrace})) is used. The horizontal lines corresponds to twice the system entropy $2H\left(\mathcal{S}\right)$, where $H(\mathcal{S})$ is calculated from $\rho_\mathcal{S} = \tr_\mathcal{E}\left[\rho_{\mathcal{SE}}\right]$. Note that $2H(\mathcal{S})$ is the maximum mutual information possible if the full system-environment state is pure as in \textbf{(a) and (c)}. For \textbf{(b) and (d)}, the mixedness of a thermal state environment reduces the maximum possible mutual information. The inset in \textbf{(a)} shows the normalised mutual information,  $I\left(\mathcal{S}:\mathcal{F}\right)/H(\mathcal{S})$.   \label{fig:quantum_darwinism}}
\end{figure*}

By applying the two partial trace methods, we are able to partition a single environment into fractions, which allows us to calculate the mutual information between system and fraction and thus determine whether or not quantum Darwinism has emerged. The results are given in Fig. \ref{fig:quantum_darwinism}, where we plot the mutual information $I\left(\mathcal{S}:\mathcal{F}\right)$ over different fractions $f$ of the environment at various times $t=300, 400, 500$ $(1/\Delta E)$. The system entropy $H(\mathcal{S})$ is calculated using the true system state $\rho_\mathcal{S} = \tr_\mathcal{E}\left[\rho_{\mathcal{SE}}\right]$, regardless of the partial trace method used to recover the system-fragment states. In Appendix~\ref{app:false-reduced-system} we recreate relevant figures using the reduced system state derived from partial trace of the system-fragment, $\rho_\mathcal{S}^\prime = \tr_\mathcal{F}\left[\rho_{\mathcal{SF}}\right]$.

As the subenvironments are not identical, the mutual information has been averaged over all possible fractions of equal size $\left|\mathcal{F}\right|$. While the mutual information value changes over different times, the mutual information \emph{relative} to the system entropy $H(\mathcal{S})$ remains roughly constant at the times considered: this is illustrated by the inset in Fig.~\ref{fig:quantum_darwinism}\textbf{(a)} which shows the normalised mutual information $I\left(\mathcal{S}:\mathcal{F}\right)/H\left(\mathcal{S}\right)$ (similar plots can be produced for the other cases---not shown here).

The shape of the plots differ between the P\'{e}rez trace and the staircase trace. Whilst the staircase trace produce mutual information plots that are typically expected in quantum Darwinism (\emph{i.e.} Fig.~\ref{fig:quantum_darwinism}\textbf{(c)} is symmetric about $f=0.5$ and about $I\left(\mathcal{S}:\mathcal{F}\right)=H(\mathcal{S})$), the P\'{e}rez trace produces a curve that quickly increases to $I\left(\mathcal{S}:\mathcal{F}\right)=H(\mathcal{S})$ before plateauing towards its maximum value (Fig. \ref{fig:quantum_darwinism} \textbf{(a)}. If the initial environment is in the thermal state (Fig.~\ref{fig:quantum_darwinism}\textbf{(b, d)}), then the maximum mutual information $I\left(\mathcal{S}:\mathcal{E}\right)\neq2H\left(\mathcal{S}\right)$, but the overall shape remains qualitatively the same as the pure initial environment. This asymmetric form is due to the environment decomposition via the P\'{e}rez trace into a direct sum $\mathcal{F}\oplus \mathcal{E}/\mathcal{F}$ \cite{Perez2010}.

No typical ``classical plateau" emerges in the staircase trace plots, and it takes a large fraction of the environment (in fact, the entire environment) to reach the quantum Darwinism condition of $I(\mathcal{S}:\mathcal{F})=H(\mathcal{S})$. Therefore we conclude that quantum Darwinism has not emerged, for neither of the initial environment states considered.

In contrast, the mutual information achieves $I\left(\mathcal{S}:\mathcal{F}\right)=H(\mathcal{S})$ for small fraction sizes with the P\'{e}rez trace. Even though Fig.~\ref{fig:quantum_darwinism}\textbf{(a)} does not match the typical plots of quantum Darwinism, it satisfies the core underlying \emph{mathematical} condition of quantum Darwinism---that of the shared information achieves $I\left(\mathcal{S}:\mathcal{F}\right)=H(\mathcal{S})$ \emph{for sufficiently small fragments} $\mathcal{F}$. Therefore, despite the non-standard curve, these results contrarily suggest that quantum Darwinism \emph{has} emerged. The plot plateaus to $I(\mathcal{S}:\mathcal{F})=2H(\mathcal{S})$, implying that larger fragments contain the \emph{full} information about the system, including quantum correlations. This does not detract from the conclusion of emergent quantum Darwinism, since it still remains true that a single fragment contains information of $H(\mathcal{S})$.

The non-symmetric nature of the plots in Fig. \ref{fig:quantum_darwinism}\textbf{(a) and (c)} is due to the properties of the P\'{e}rez trace. As we mentioned in subsection \ref{subsubsec:PerezTrace} and further in Appendix~\ref{app:false-reduced-system}, the P\'erez trace, although physically motivated, has a caveat: \emph{i.e.}, it does not give the correct system state from the reduced system-fragment state: $\rho_\mathcal{S}\neq \tr_\mathcal{F} \rho_\mathcal{SF}$.
 The authors of Ref.~\cite{Blume-Kohout2005} show that mutual information plots should be symmetric, under the assumption of a partial trace that gives the correct reduced system state. The proof relies on the fact that $\rho_\mathcal{SE}$ is a pure state. By partitioning the environment $\mathcal{E}$ into exactly two parts, $\mathcal{E}_1$ and $\mathcal{E}_2$, then $I(\mathcal{S}:\mathcal{E}_1)+I(\mathcal{S}:\mathcal{E}_2)=I(\mathcal{S}:\mathcal{E}_1\mathcal{E}_2)$. This is because for pure $\rho_\mathcal{SE}=\rho_{\mathcal{SE}_1\mathcal{E}_2}$, we have that $H(\mathcal{S})=H(\mathcal{E}_1\mathcal{E}_2)$, $H(\mathcal{E}_1)=H(\mathcal{S}\mathcal{E}_2)$, and $H(\mathcal{E}_2)=H(\mathcal{S}\mathcal{E}_1)$ (since bipartitions of pure states have the same spectrum). For the P\'{e}rez trace however, the reduced states $\rho_{\mathcal{S}\mathcal{E}_2}$ and $\rho_{\mathcal{S}\mathcal{E}_1}$ are pure if $\rho_{\mathcal{SE}_1\mathcal{E}_2}$ is pure, however $\rho_{\mathcal{E}_1}$ and $\rho_{\mathcal{E}_2}$ are not pure in general, leading to $H(\mathcal{E}_1)\neq H(\mathcal{S}\mathcal{E}_2)$,$H(\mathcal{E}_2)\neq H(\mathcal{S}\mathcal{E}_1)$, and in fact $I(\mathcal{S}:\mathcal{E}_1)+I(\mathcal{S}:\mathcal{E}_2)\geq I(\mathcal{S}:\mathcal{E}_1\mathcal{E}_2)$. Therefore, the mutual information plots are not symmetric in contrast with Ref.~\cite{Blume-Kohout2005}.

Furthermore, the sharp mutual information rise in  Fig.~\ref{fig:quantum_darwinism}\textbf{(a)} is in fact a \emph{universal feature} of the P\'{e}rez trace when the full system-environment state is pure---we \emph{always} have $I\left(\mathcal{S}:\mathcal{F}\right)=H\left(\mathcal{S}\right)$ when $\left|\mathcal{F}\right|=1$, and $I\left(\mathcal{S}:\mathcal{F}\right)>H\left(\mathcal{S}\right)$ for fragment sizes $\left|\mathcal{F}\right|>1$. This is because the partially reduced $\mathcal{SF}$ state is always effectively pure or zero, in which case $H\left(\mathcal{SF}\right)=0$ always for $\mathcal{F}\neq\emptyset$. Furthermore, when $\mathcal{F}$ consists of a \emph{single} level, then $\rho_{\mathcal{F}}$ is essentially a c-number, and due to trace preservation, is either $\rho_{\mathcal{F}}=\left[1\right]$ or $\rho_{\mathcal{F}}=\left[0\right]$, and so $H\left(\mathcal{F}\right)=0$ for $\left|\mathcal{F}\right|=1$.

Hence, in the scenario where the environment structure implicitly only allows us access to specific levels, quantum Darwinism either (1) always emerges or (2) is not applicable to this scenario. The results to follow later in this paper conflict with the first conclusion. If we take the second conclusion, then this shows that quantum Darwinism is limited in its applicability and cannot be used to universally explore the emergence of objectivity. Thus, regardless, these results and further analyses presented later in the paper show that quantum Darwinism\textemdash given by the mathematical condition $I(\mathcal{S}:\mathcal{F})=H(\mathcal{S})$ for sufficiently small $\mathcal{F}$\textemdash is inconsistent.

\subsection{The quantum versus classical nature of the mutual information}

The mutual information of two systems can be decomposed into the sum of their accessible (Holevo) information $\chi$ and the quantum discord $\mathcal{D}$ between them \cite{Zwolak2013}:
\begin{equation}
I\left(\mathcal{S}:\mathcal{F}\right)=\chi\left(\mathcal{S}:\mathcal{F}\right)+\mathcal{D}\left(\mathcal{S}:\mathcal{F}\right).
\end{equation}
The accessible information quantifies about the amount of classical information shared, whereas the discord describes any non-classical correlations \cite{Ollivier2001,Henderson2001}. When the mutual information between system and fragment is approximately equal to the system entropy, $I\left(\mathcal{S}:\mathcal{F}\right)\approx H\left(\mathcal{S}\right)$, quantum Darwinism posits that the fragment $\mathcal{F}$ contains (approximately) all the information about the properties of the system. \citet{Brandao2015} prove that for sufficiently large environments, the mutual information comprises mostly of the accessible ``classical'' information $I\left(\mathcal{S}:\mathcal{F}\right)\approx\chi\left(\mathcal{S}:\mathcal{F}\right)$ (optimised over POVMs), and this fact has been used to derive estimations of the redundancy based on the accessible information \cite{Zwolak2016,Zwolak2014,Zwolak2013}. In this study of emerging classicality, it is reasonable to expect that an objective state would have a large amount of accessible information.

However, a recent work \cite{Pleasance2017} suggests that this is not always the case---they show a scenario whereby there is apparent quantum Darwinism due to the present mutual information plateau, but where the mutual information, at small fractions, is largely comprised of \emph{quantum discord}. As such, it is imperative for us to investigate that, considering that P\'{e}rez trace suggested emergent quantum Darwinism in conflict with the staircase trace.

We calculate the accessible information by maximising over possible POVMs on the system:
\begin{align}
\chi\left(\mathcal{S}:\mathcal{F}\right) & =\max_{\hat{\Pi}_{\mathcal{S}}}\left\{\begin{array}{l}	
     H\left(\sum_{a}p\left(a\right)\rho_{\mathcal{F}|a}\right) \\
     -\sum_{a}p\left(a\right)H\left(\rho_{\mathcal{F}|a}\right)
\end{array} \right\},
\end{align}
where $a$ are the measurement results of the POVM $\hat{\Pi}_{\mathcal{S}}$, $\rho_{\mathcal{F}|a}$ refers to the state of the fragment given result $a$ and $p\left(a\right)$ is the probability of result $a$ \cite{Zwolak2013}. The discord can always be minimised using rank one projectors \cite{Datta2008phd}, and so corresponding the accessible information can always be maximised using rank one projectors. As our system is two-level, a general observable can be written in form $\vec{r}\cdot\vec{\sigma}$, where $\vec{\sigma}=\left(\sigma_{x},\sigma_{y},\sigma_{z}\right)$, and $\vec{r}$ is a unit vector. The two associated projectors are then $\hat{\Pi}_{\mathcal{S}}^{\pm}=\left(\id\pm\vec{r}\cdot\vec{\sigma}\right)/2$ and the conditional fragment states are 
\begin{align}
\rho_{\mathcal{F}|\pm} & =\tr_{\mathcal{S}}\left(\hat{\Pi}_{\mathcal{S}}^{\pm}\rho_{\mathcal{SF}}\hat{\Pi}_{\mathcal{S}}^{\pm}\right)/p\left(\pm\right),
\end{align}
where $p\left(\pm\right)=\tr\left(\rho_{\mathcal{SF}}\hat{\Pi}_{\mathcal{S}}^{\pm}\right)$. We optimise via a random search, where the unit vector $\vec{r}$ is randomly chosen across the unit-sphere.

\begin{figure*}
\begin{centering}
\includegraphics[width=1\textwidth]{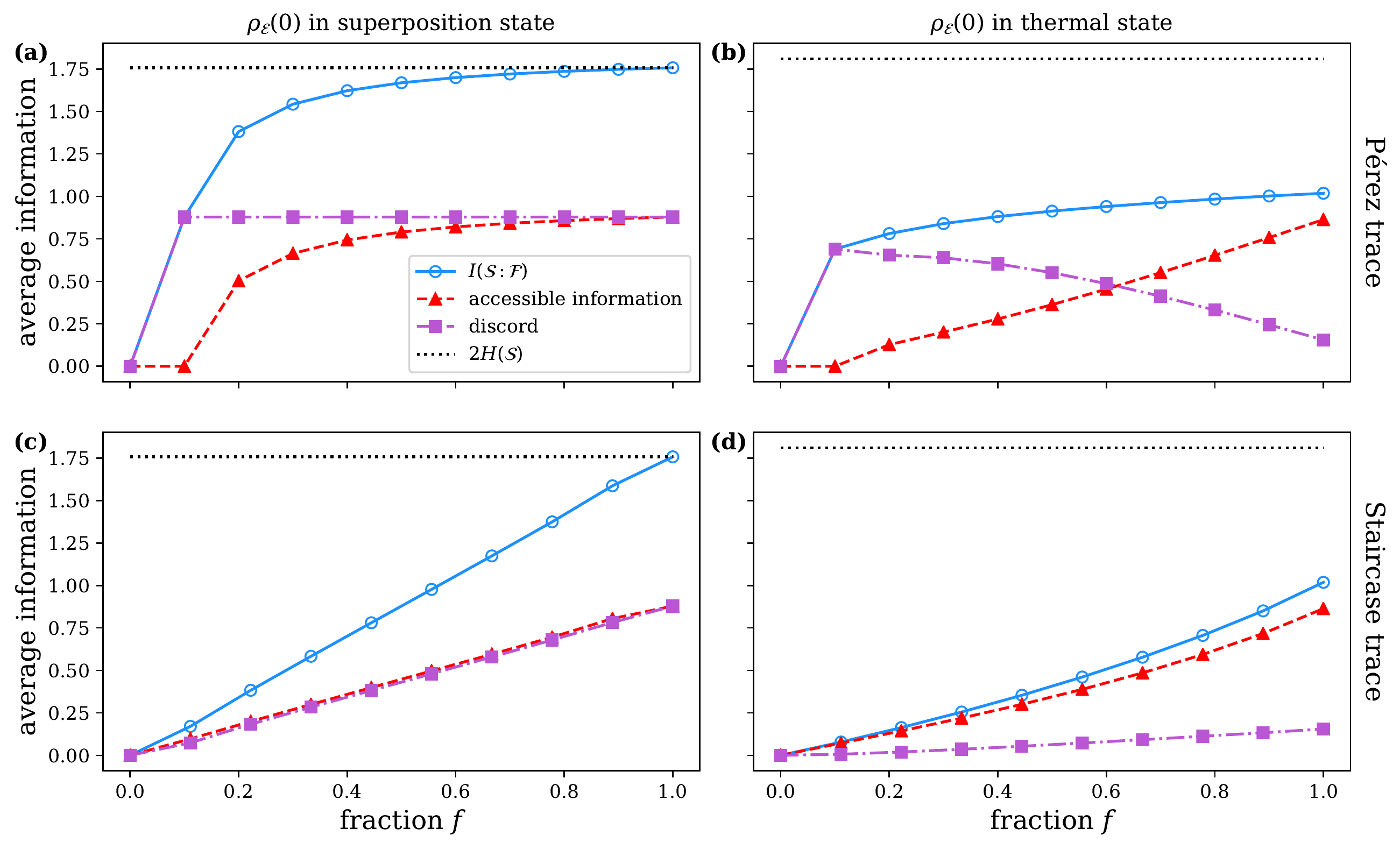}
\par\end{centering}
\caption{\textbf{Decomposition of mutual information $I(\mathcal{S}:\mathcal{F})$ into accessible information and quantum discord (at time $t=500$)} The accessible information and quantum discord is averaged over all environment fractions of the same size $f$. The system entropy is $H(\mathcal{S})$. For all figures, the parameters are $\Delta E=1$, $\delta\varepsilon=\Delta E$, $\lambda=\Delta E/5$, and $\beta=10$ for the thermal state ($\hbar=1$). \label{fig:accessible_information}}
\end{figure*}

The results are given in Fig.~\ref{fig:accessible_information}. We find that the amount of quantum discord \emph{is comparable} to the amount of accessible information at the various different fraction sizes $f$. A relatively \emph{large} fraction of the environment is required to obtain (accessible) information approximately equal to the system entropy. This is true regardless of partial trace method and regardless of initial environment state. By decomposing the mutual information into its ``classical'' and ``quantum'' components, the conclusion is clear: the system state is \emph{not} objective, regardless of what quantum Darwinism would otherwise imply.

\subsection{Summary}

In this section, we described two methods to partition a single environment into fractions. By doing so, we are able to evaluate the mutual information between system and fraction to determine whether or not quantum Darwinism emerged. Both methods have been shown to be viable, opening up many more environments that could be studied. The caveats to the methods is that the level-partitioning and elimination P\'{e}rez trace does not satisfy all the usual properties of the partial trace; and the staircase trace assumes a particular structured environment. Notwithstanding, the spirit of the staircase trace can be used to produce yet more alternative trace methods based on assumed environment structure.

The different partial traces came to different conclusions about whether or not quantum Darwinism emerged, whereby quantum Darwinism is supposed to emerge when $I(\mathcal{S}:\mathcal{F})\approx H(\mathcal{S})$ is achieved with small fragments $\mathcal{F}$. 
However, \emph{both} the methods agreed that the mutual information is \emph{not} comprised mostly of ``classical'' accessible information, but rather a roughly equal mixture of accessible information and quantum discord. From this point of view, both the methods conclude that the system is \emph{not} objective, regardless of apparent quantum Darwinism. Granted, the environment comprised of $N$ or $N-1$ artificial subsystems, while the majority of the general results on quantum Darwinism requires the large environment limit. However, our results have shown that quantum Darwinism is either inconsistent or not sufficiently applicable in general in determining whether or not the quantum-to-classical transition occurred. As such, we consider a more stringent condition on objectivity: spectrum broadcast structure.

\section{Spectrum Broadcasting \label{sec:Spectrum-Broadcast-Structure}}

A system $\mathcal{S}$ and a collection of sub-environments $\mathcal{E}_{1},\ldots,\mathcal{E}_{fN}$ (that form the fragment $\mathcal{F}$) is said have spectrum broadcast structure when it can be written in the form 
\begin{align}
\rho_{\mathcal{SF}} & =\sum_{i}p_{i}\ket{i}_{S}\bra{i}\otimes\rho_{i}^{\mathcal{E}_{1}}\otimes\cdots\otimes\rho_{i}^{\mathcal{E}_{fN}},\label{eq:SBS}
\end{align}
where $\left\{ \ket{i}\right\} $ is the basis of the pointer states in the system's space, $p_{i}$ are probabilities, and all states $\rho_{i}^{\mathcal{E}_{k}}$ are perfectly distinguishable: $\rho_{i}^{\mathcal{E}_{k}}\rho_{j}^{\mathcal{E}_{k}}=0$ for all $i\neq j$ \cite{Horodecki2015}. We calculate an upperbound of the minimum distance between the true system-fragment state $\rho_{\mathcal{SF}}$ and ideal (and unknown) nearest spectrum broadcast structure state $\rho_{\mathcal{SF}}^{SBS}$ by adapting the bound derived in Ref.~\cite{Mironowicz2017}. By doing so, it is clear that spectrum broadcasting has not occurred, confirming the non-objectivity conclusion from the previous section.

\subsection{An approximate measure of spectrum broadcasting}

One possible measure of spectrum broadcasting structure in a system-fragment state is
\begin{align}
D_{SBS}\left(\rho_{\mathcal{SF}}\right) & =\min_{\rho_{\mathcal{SF}}^{\text{SBS}}}\bigl\Vert\rho_{\mathcal{SF}}-\rho_{\mathcal{SF}}^{\text{SBS}}\bigr\Vert,
\end{align}
minimised over all states of spectrum-broadcast form, under the $L1$ norm. However, this suffers all the analogous difficulties in measuring the distance of entangled states to the set of separable states---the spectrum-broadcast states are a subset of the separable states and brute-force optimisation is no easy task.

\citet{Mironowicz2017} take the approach of constructing a computable error bound $\eta\left[\rho_{\mathcal{SF}}\right]$ to the distance given by the summation of the decoherence factors and the distinguishability of fraction states $\rho_{i}^{\mathcal{E}_{k}}$ under the case of the quantum measurement limit. Using the same derivation (see Appendix \ref{app:Upper-bound-the-minimum} and Ref. \cite{Mironowicz2017}), we rewrite that bound in a more general form suitable for our model:
\begin{align}
D_{SBS}\left(\rho_{\mathcal{SF}}\right) & \leq\eta\left[\rho_{\mathcal{SF}}\right],  \\
\eta\left[\rho_{\mathcal{SF}}\right] &\equiv \min_{\left\{ P_{i}^{\mathcal{S}}\right\} }\left[
\begin{array}{l}	
     \bigl\Vert\rho_{\mathcal{SF}}-\rho_{\mathcal{SF}}^{\text{sep}}\left(\left\{ P_{i}^{\mathcal{S}}\right\} \right)\bigr\Vert_{1} \\
     +\sum_{i\neq j}\sqrt{p_{i}p_{j}}B\left(\rho_{i}^{\mathcal{F}},\rho_{j}^{\mathcal{F}}\right)
\end{array} \right].\label{eq:SBS_distance}
\end{align}
where $B\left(\rho_{1},\rho_{2}\right)=\bigl\Vert\sqrt{\rho_{1}}\sqrt{\rho_{2}}\bigr\Vert_{1}$
is the fidelity, and where the set of rank-one \emph{system} projectors $\left\{ P_{i}^{\mathcal{S}}=\ket{i'}\bra{i'}\right\} $ allows us to construct the separable state
\begin{align}
\rho_{\mathcal{SF}}^{\text{sep}}\left(\left\{ P_{i}^{\mathcal{S}}\right\} \right) &=\sum_{i}P_{i}^{\mathcal{S}}\otimes\id^{\mathcal{F}}\rho_{\mathcal{SF}}P_{i}^{\mathcal{S}}\otimes\id^{\mathcal{F}}\\
&\eqqcolon\sum_{i=0,1}p_{i}\ket{i^{\prime}}\bra{i^{\prime}}\otimes\rho_{i}^{\mathcal{F}},
\end{align}
where $p_i = \tr\left[P_{i}^{\mathcal{S}}\rho_{\mathcal{SF}}\right]$ and $\rho_{i}^{\mathcal{F}} =\tr_\mathcal{S}\left[P_{i}^{\mathcal{S}}\rho_{\mathcal{SF}}\right]/p_i$.
The set of projectors is optimised over to minimise the error bound $\eta[\rho_\mathcal{SF}]$. This amounts to talking an optimal instantaneous pointer basis $\{\ket{i^\prime}\bra{i^\prime}\}_i$ upon which the shared system-environment information is maximised and $\eta\left[\rho_{\mathcal{SF}}\right]$ is minimised. The bound $\eta\left[\rho_{\mathcal{SF}}\right]$ is tight when $\rho_{\mathcal{SF}}$ has spectrum broadcast structure. For further details, see Appendix \ref{app:Upper-bound-the-minimum}.

\subsection{Results}

\begin{figure*}
\begin{centering}
\includegraphics[width=1\textwidth]{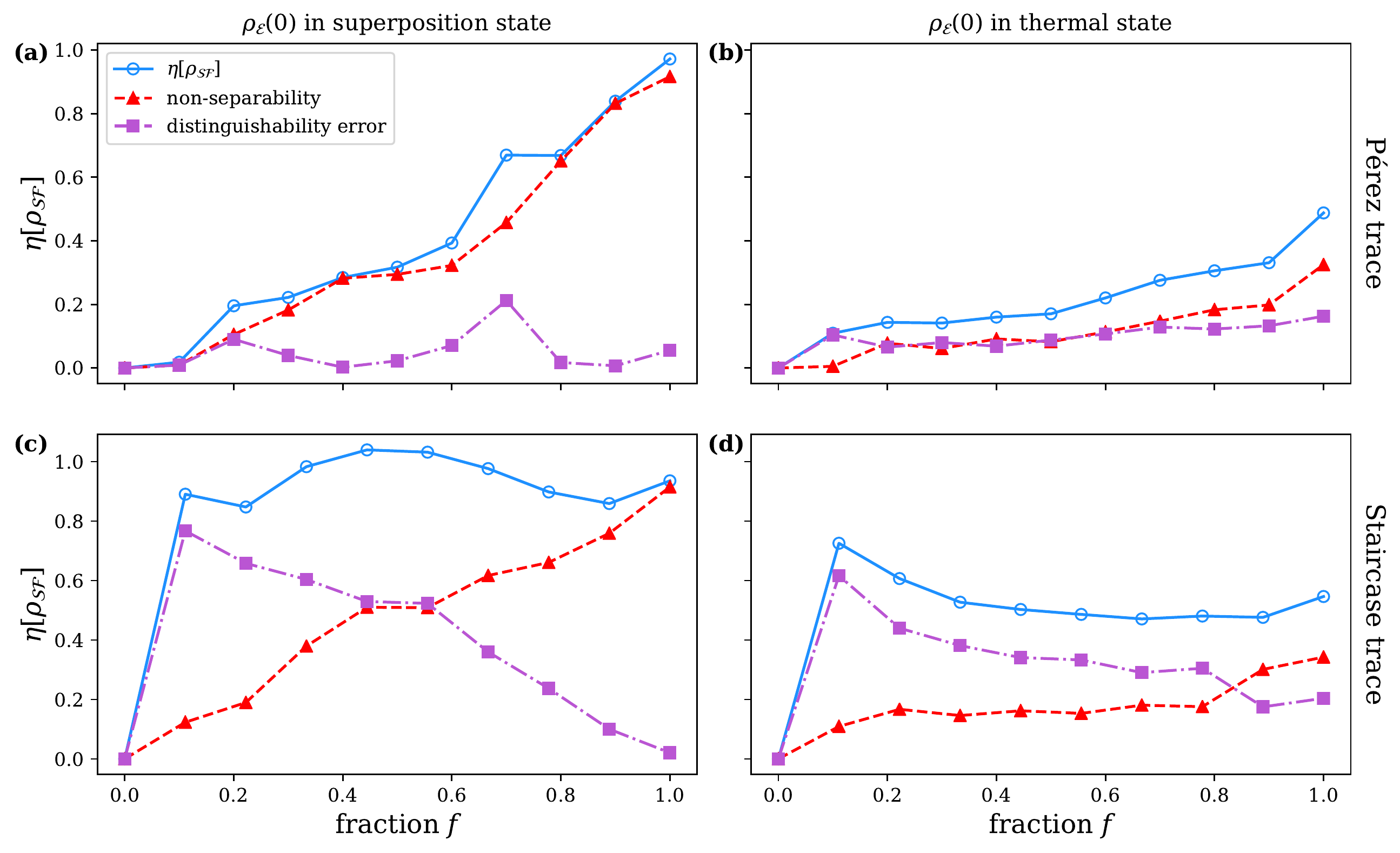}
\par\end{centering}
\caption{\textbf{Bounds to the distance of system-fragment states to the set of spectrum broadcast structure states (at time $t=500$).} The distance $D_{SBS}\left(\rho_{\mathcal{SF}}\right)$ is bounded by $\eta\left[\rho_{\mathcal{SF}}\right]$, the summation of the non-separability of the system-fragment state (first term in Eq. (\ref{eq:SBS_distance})) and the distinguishability error of the reduced fragment states (second term in Eq. (\ref{eq:SBS_distance})). Overall, the system-fragment states do not have a spectrum broadcast structure since $\eta\left[\rho_{\mathcal{SF}}\right]\neq 0$. For all figures, the parameters are $\Delta E=1$, $\delta\varepsilon=\Delta E$, $\lambda=\Delta E/5$, and $\beta=10$ for the thermal state ($\hbar=1$). \label{fig:SBS_breakdown}}
\end{figure*}

Since the system is a qubit, the projectors can be written as $P_{\pm}^{\mathcal{S}}=\left(\id\pm\hat{n}\cdot\vec{\sigma}\right)/2$ for unit vector $\hat{n}$. Numerically, we minimise the distance given in Eq. (\ref{eq:SBS_distance}) by sampling $\hat{n}$ over the unit sphere. We further minimise over all fractions of the same size, to produced Fig. \ref{fig:SBS_breakdown}.

We find that there is \emph{no} fraction size at which a spectrum broadcast structure is formed between system and environment fragment in almost all cases considered. This confirms our prior conclusion that objectivity has not emerged. The only ``prospective'' case is that of the smallest non-zero environment fraction in the P\'{e}rez trace (Fig. \ref{fig:SBS_breakdown}(a)), where the distance is vanishing. At this point however, the environment-fragment consists of a single c-number $\rho_{\mathcal{F}}=\left[1\right]$ or $\rho_{\mathcal{F}}=\left[0\right]$, hence is not a true spectrum broadcast structure.

Interestingly, the ``spectrum-broadcastness'' of the system-environment state differs between the partial trace methods used: The P\'{e}rez trace finds that the system-fragment state tends to be non-separable, whereas the staircase trace assumptions find that for small fragments, the states are largely separable but rather non-distinguishable. We attribute this difference to the differing assumptions of the structure of the environment implicit in either partial trace. 

\subsection{Summary}

By using a manageable upper-bound on the distance of a state to spectrum broadcast form, we investigated the structure of the system-fragment states. Both partial traces agreed that the system-fragment states did \emph{not} have spectrum broadcast structure form, strengthening our prior conclusion that objectivity has not emerged. The minimisation could, in future, be done via more sophisticated methods than the random search employed here. Nonetheless, our work illustrates its practicality, and the combined investigation of both quantum Darwinism \emph{and} spectrum broadcasting leads to stronger conclusions on the question of emergent classicality.

\section{Conclusions \label{sec:Conclusions}}

We jointly investigated quantum Darwinism and spectrum broadcasting as frameworks to test the objectivity of a two-level system interacting with an effective $N$-level environment via a random matrix interaction. We compared two methods to break the environment into fragments: the partial trace of \citet{Perez2010}, which partitions the levels and eliminates any components of levels outside the relevant fraction, and our proposed staircase partial trace. The former has the caveat that system density matrix can only be truly recovered from the trace of the full environment $\rho_{\mathcal{SE}}$ but not the fraction $\rho_{\mathcal{SF}}$, whilst our proposed method does not have this problem, instead assuming that the environment is structured such that each increasing energy level is the excited level of a single particular subenvironment.

Using a numerical distance measure, we found that the system-fragment states did not have spectrum broadcast structure. We also found that the mutual information (used to evaluate quantum Darwinism) is comprised of comparable amounts of both ``classical'' accessible information \emph{and} quantum discord.
However, since quantum Darwinism is the evaluated using the quantum mutual information---an indiscriminate sum of accessible information and quantum discord---we found a case whereby it appeared as though quantum Darwinism had emerged\textemdash \emph{i.e.} when using the  partial trace of \citet{Perez2010}.

By exploring the decomposition of the mutual information into accessible information and quantum discord, we have been able to illustrate a discrepancy between quantum Darwinism and supposed state objectivity under the P\'{e}rez trace: a ``mutual information plateau'' and the shared information condition $I(\mathcal{S}:\mathcal{F})=H(\mathcal{S})$ are \emph{not} sufficient for state objectivity. Combined with the results of Ref.~\cite{Pleasance2017}, this suggests that quantum Darwinism could potentially emerge even if the mutual information is largely quantum in nature. Unlike previous work, we have also compared the results of quantum Darwinism with that of spectrum broadcasting. Contrary to quantum Darwinism, the investigation of state structure does not give any false positives on the emergence of objectivity, regardless of the partial tracing method used. In contrast, we have shown that quantum Darwinism, when investigated with the P\'erez trace, is inconsistent with spectrum broadcast structure. This suggests that the formalism of quantum Darwinism may be inconsistent in general, or that quantum Darwinism has limited applicability to only certain types of environments. We suggest that future studies of quantum Darwinism should take into account the amount of quantum discord, and our results demonstrate that state structure analyses are feasible. In highlighting the discrepancy between quantum Darwinism and spectrum broadcasting---via the potential failure of quantum Darwinism in regards to the nature of the quantum mutual information---our work opens the door to investigate the general conditions under which quantum Darwinism (or a modified version of it) would be exactly equivalent to spectrum broadcast structure.

We have shown that it is possible to study (apparently) monolithic environments within the quantum Darwinism and spectrum broadcasting frameworks by using suitable methods to partition the environment. We considered two different partial trace methods in this paper: the level-partitioning and elimination P\'erez trace and the staircase environment trace. The novel staircase environment trace we introduce has the advantage over the P\'erez trace in that it commutes with the traditional partial trace \emph{and} it produces traditional mutual information plots, allowing the quantum Darwinism formalism to be applied. The staircase environment trace assumed a particular type of structured environment: the spirit of trace construction can be applied to other types of structured environments. For example, one can also construct a partial trace that assumes the environment is composed of $N-1$ identical two-level systems, with increasing environment energy corresponding to consecutively exciting the different subsystems.

The quantum-to-classical transition is a fundamental issue in quantum mechanics. Within the frameworks of quantum Darwinism and spectrum broadcasting, the correlations between the system and environment are studied for various different markers of state objectivity. The environment needs to be decomposable: for an environment of $N$ subsystems, this would correspond to a Hilbert space of at least $2^{N}$. The alternative partial trace methods introduced in this paper make it possible to study environments of $N$ subsystems with a (drastically reduced) Hilbert space of $N$ dimensions. This work thus opens up classes of environments that can now be studied using the tools of quantum Darwinism and spectrum broadcast structure. 

\section*{Acknowledgments}

We thank the anonymous referee for constructive feedback on previous versions of this manuscript. This work was supported by the Engineering and Physical Sciences Research Council {[}grant number EP/L015242/1{]}.

\bibliographystyle{apsrev4-1}
\bibliography{main}

\appendix

\section{Bounding the distance to the set of spectrum broadcast structure states\label{app:Upper-bound-the-minimum}}

The following derivation follows that of \citet{Mironowicz2017}, with the notation adjusted to match the main text of this paper.

Consider writing $\rho_{\mathcal{SF}}$ in some particular basis $\left\{ \ket{i}\right\} $ (which need not be the standard computation basis):
\begin{align}
\rho_{\mathcal{SF}} & =\sum_{i,j=0,1}\sum_{n,m\in\mathcal{F}}c_{ijnm}\ket{i}\bra{j}\otimes\ket{n}\bra{m}\\
 & \equiv\rho_{\mathcal{SF}}^{\text{sep}}+\sigma_{\mathcal{SF}},
\end{align}
where we have defined the separable component of the system-fragment
state
\begin{align}
\rho_{\mathcal{SF}}^{\text{sep}} & \equiv\sum_{i=0,1}\sum_{n,m\in\mathcal{F}}c_{iinm}\ket{i}\bra{i}\otimes\ket{n}\bra{m}\\
 & =\sum_{i=0,1}p_{i}\ket{i}\bra{i}\otimes\rho_{i}^{\mathcal{F}},
\end{align}
with $\rho_{i}^{\mathcal{F}}=\left(1/p_{i}\right)\sum_{n,m\in\mathcal{F}}c_{iinm}\ket{n}\bra{m}$
and $p_{i}=\sum_{n\in\mathcal{F}}c_{iinn}$. The non-separable remainder
is
\begin{align}
\sigma_{\mathcal{SF}} & \equiv\sum_{i\neq j}\sum_{n,m\in\mathcal{F}}c_{ijnm}\ket{i}\bra{j}\otimes\ket{n}\bra{m}.
\end{align}

Suppose we employ the following complete set of projectors $\left\{ \Pi_{i}^{\mathcal{F}}\right\} _{i}$ to measure the environment fragment and attempt to discriminate between $\rho_{i}^{\mathcal{F}}$ states (which we will later optimise over). A projection of $\Pi_{i}$ will result in state $\Pi_{i}^{\mathcal{F}}\rho_{i}^{\mathcal{F}}\Pi_{i}^{\mathcal{F}}$ with probability $\tr\left[\rho_{i}^{\mathcal{F}}\Pi_{i}^{\mathcal{F}}\right]$. With probability $1-\tr\left[\rho_{i}^{\mathcal{F}}\Pi_{i}^{\mathcal{F}}\right]=\tr\left[\rho_{i}^{\mathcal{F}}\left(\id^{\mathcal{}}-\Pi_{i}^{\mathcal{F}}\right)\right]$,
we will obtain an error. The error in discriminating between the different $\rho_{i}^{\mathcal{F}}$ states (which occur with probability $p_{i}$) is then \cite{Mironowicz2017} 
\begin{align}
\text{Err}\left[\left\{ p_{i},\rho_{i}^{\mathcal{F}}\right\} ,\left\{ \Pi_{i}^{\mathcal{F}}\right\} \right] & =\sum_{i}p_{i}\tr\left[\rho_{i}^{\mathcal{F}}\left(\id-\Pi_{i}^{\mathcal{F}}\right)\right].
\end{align}
The ideal spectrum broadcast structure state for the set of measurements
$\left\{ \Pi_{i}^{\mathcal{F}}\right\} _{i}$ is 
\begin{align}
\rho_{\mathcal{SF}}^{\text{SBS}} & =\sum_{i}q_{i}\ket{i}\bra{i}\otimes\tilde{\rho}_{i}^{\mathcal{F}},
\end{align}
where $\tilde{\rho}_{i}^{\mathcal{F}}=\Pi_{i}^{\mathcal{F}}\rho_{i}^{\mathcal{F}}\Pi_{i}^{\mathcal{F}}/\tr\left[\rho_{i}^{\mathcal{F}}\Pi_{i}^{\mathcal{F}}\right]$
are states that are now perfectly distinguishable by $\left\{ \Pi_{i}^{\mathcal{F}}\right\} _{i}$,
and $q_{i}=p_{i}\tr\left[\rho_{i}^{\mathcal{F}}\Pi_{i}^{\mathcal{F}}\right]/\sum_{j}p_{j}\tr\left[\rho_{j}^{\mathcal{F}}\Pi_{i}^{\mathcal{F}}\right]$
are the normalised probabilities.

Using this candidate form of a possible spectrum broadcast structure state,
\begin{align}
D_{SBS}\left(\rho_{\mathcal{SF}}\right) & \leq\bigl\Vert\rho_{\mathcal{SF}}-\rho_{\mathcal{SF}}^{\text{SBS}}\bigr\Vert_{1} \\
 &=\bigl\Vert\rho_{\mathcal{SF}}^{\text{sep}}+\sigma_{\mathcal{SF}}-\rho_{\mathcal{SF}}^{\text{SBS}}\bigr\Vert_{1}\\
 & \leq\bigl\Vert\rho_{\mathcal{SF}}^{\text{sep}}-\rho_{\mathcal{SF}}^{\text{SBS}}\bigr\Vert_{1}+\bigl\Vert\sigma_{\mathcal{SF}}\bigr\Vert_{1},
\end{align}
following from the triangle inequality. The second term $\bigl\Vert\sigma_{\mathcal{SF}}\bigr\Vert_{1}=2T\left(\rho_{\mathcal{SF}},\rho_{\mathcal{SF}}^{\text{sep}}\right)$ can be easily calculated numerically. Focusing on the first term,
\begin{multline}
\bigl\Vert\rho_{\mathcal{SF}}^{\text{sep}}-\rho_{\mathcal{SF}}^{\text{SBS}}\bigr\Vert_{1}
\leq\sum_{i=0,1}p_{i}\left\|\rho_{i}^{\mathcal{F}}-\dfrac{\Pi_{i}\rho_{i}^{\mathcal{F}}\Pi_{i}}{\sum_{j}p_{j}\tr\left[\rho_{j}^{\mathcal{F}}\Pi_{j}\right]}\right\|_{1},
\end{multline}
which follows from the triangle inequality, and that $\bigl\Vert\ket{i}\bra{i}\otimes F_{i}\bigr\Vert_{1}=\tr\left[\ket{i}\bra{i}\right]\bigl\Vert F_{i}\bigr\Vert_{1}$. Since
\begin{align}
\left\|\rho_{i}^{\mathcal{F}}-\dfrac{\Pi_{i}\rho_{i}^{\mathcal{F}}\Pi_{i}}{\sum_{j}p_{j}\tr\left[\rho_{j}^{\mathcal{F}}\Pi_{j}\right]}\right\|_{1} & =\tr\left[\rho_{i}^{\mathcal{F}}-\dfrac{\Pi_{i}\rho_{i}^{\mathcal{F}}\Pi_{i}}{\sum_{j}p_{j}\tr\left[\rho_{j}^{\mathcal{F}}\Pi_{j}\right]}\right] \nonumber \\
 & \leq\tr\left[\rho_{i}^{\mathcal{F}}\left(\id-\Pi_{i}\right)\right],
\end{align}
we have that $\bigl\Vert\rho_{\mathcal{SF}}^{\text{sep}}-\rho_{\mathcal{SF}}^{\text{SBS}}\bigr\Vert_{1}\leq\text{Err}\left[\left\{ p_{i},\rho_{i}^{\mathcal{F}}\right\} ,\left\{ \Pi_{i}^{\mathcal{F}}\right\} \right],$
which, after minimisation, is bounded by the optimal discrimination error:
\begin{align}
\min_{\rho_{\mathcal{SF}}^{\text{SBS}}}\bigl\Vert\rho_{\mathcal{SF}}^{\text{sep}}-\rho_{\mathcal{SF}}^{\text{SBS}}\bigr\Vert_{1} & \leq\min_{\left\{ \Pi_{i}^{\mathcal{F}}\right\} }\text{Err}\left[\left\{ p_{i},\rho_{i}^{\mathcal{F}}\right\} ,\left\{ \Pi_{i}^{\mathcal{F}}\right\} \right]\\
 & \leq\sum_{i\neq j}\sqrt{p_{i}p_{j}}B\left(\rho_{i}^{\mathcal{F}},\rho_{j}^{\mathcal{F}}\right),
\end{align}
where $B\left(\rho_{1},\rho_{2}\right)=\bigl\Vert\sqrt{\rho_{1}}\sqrt{\rho_{2}}\bigr\Vert_{1}$ is the fidelity \cite{Mironowicz2017}. Hence, 
\begin{align}
D_{SBS}\left(\rho_{\mathcal{SF}}\right) & \leq \min_{\sigma_{\mathcal{SF}}}\left[
\begin{array}{l}	
     \bigl\Vert\sigma_{\mathcal{SF}}\bigr\Vert_{1} \\
     +\sum_{i\neq j}\sqrt{p_{i}p_{j}}B\left(\rho_{i}^{\mathcal{F}},\rho_{j}^{\mathcal{F}}\right)
\end{array} \right],
\end{align}
where the minimisation over $\sigma_{\mathcal{SF}}$ translates to minimising over the optimal basis $\left\{ \ket{i}\right\} $ for system, \emph{i.e.},  minimising over the set of \emph{system} projectors $\left\{ P_{i}^{\mathcal{S}}\right\}$ that define $\rho_{\mathcal{SF}}^{\text{sep}}\left(\left\{ P_{i}^{\mathcal{S}}\right\} \right)=\sum_{i}P_{i}^{\mathcal{S}}\rho_{\mathcal{SF}}P_{i}^{\mathcal{S}}$ and $\sigma_{\mathcal{SF}}=\rho_{\mathcal{SF}}-\rho_{\mathcal{SF}}^{\text{sep}}\left(\left\{ P_{i}^{\mathcal{S}}\right\} \right)$.

\newpage

\section{If the reduced system is derived from the system-fragment state}\label{app:false-reduced-system}

For clarity, we repeat the definition of the P\'erez trace, which from the general system-environment state
\begin{align}
\rho_{\mathcal{SE}}\left(t\right) & =\sum_{i,j=0,1}\sum_{n,m=0}^{N-1}c_{ijnm}\ket{i}_{\mathcal{S}}\bra{j}\otimes\ket{n}_{\mathcal{E}}\bra{m},
\end{align}
produces the following reduced system-fragment state:
\begin{align}
\rho_{\mathcal{SF}}\left(t\right) & =\dfrac{1}{N_{\mathcal{F}}}\sum_{i,j=0,1}\sum_{n,m\in\mathcal{F}}c_{ijnm}\ket{i}_{\mathcal{S}}\bra{j}\otimes\ket{n}_{\mathcal{E}}\bra{m},
\end{align}
where $N_{\mathcal{F}}=\sum_{i=0,1}\sum_{n\in\mathcal{F}}c_{iinn}$ is the normalisation factor.

As noted in the main text, the true state of the system, $\rho_\mathcal{S}=\tr_\mathcal{E}\left[\rho_\mathcal{SE}\right]$, cannot be determined from the system-fragment $\rho_\mathcal{SF}$ if using the P\'erez partial trace (Section~\ref{subsubsec:PerezTrace}). As such, when we calculate the mutual information, $I(\mathcal{S}:\mathcal{F})=H(\mathcal{S})+H(\mathcal{F})-H(\mathcal{SF})$, we took $H(\mathcal{S})$ as being the true entropy of the system, determined by the true state of the system.

Arguably, we could have taken $H(\mathcal{S})$ as being determined by the system state calculated from the reduced system-environment, $\rho_\mathcal{S}'=\tr_\mathcal{F}[\rho_\mathcal{SF}]$. In this appendix, we show what happens if one had taken this route. Note that the staircase trace produces identical plots to those shown in the main text---\emph{i.e.}, the staircase trace recovers the correct system state.

\begin{figure*}
\begin{centering}
\includegraphics[width=1\textwidth]{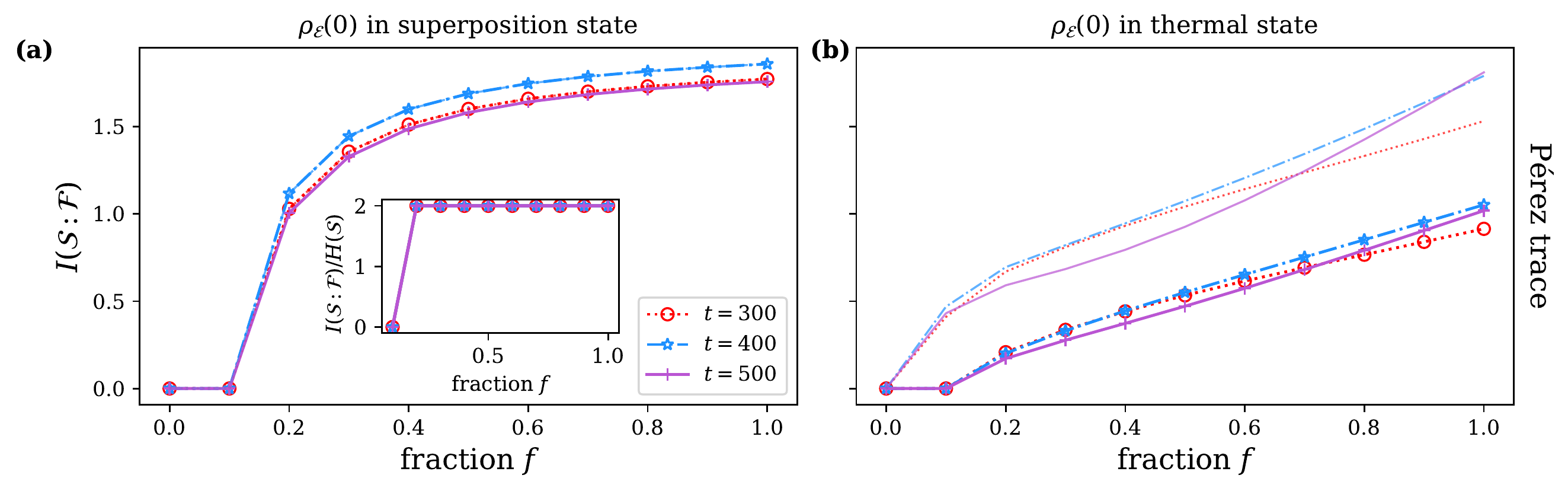}
\par\end{centering}
\caption{\textbf{Mutual information $I\left(\mathcal{S}:\mathcal{F}\right)$ between system and environment fragments using the P\'erez partial trace  (Eq. (\ref{eq:PerezTrace}))} (compare with Fig.~\ref{fig:quantum_darwinism}). The \citet{Perez2010} trace is used to form different fractions of the environment. The reduced system is calculated as $\rho_\mathcal{S}^\prime = \tr_\mathcal{F}\left[\rho_{\mathcal{SF}}\right]$. The dashed lines (without markers) corresponds to twice the system entropy $2H\left(\mathcal{S}\right)$ (the maximum mutual information possible if the full system-environment state is pure). The inset in \textbf{(a)} shows the normalised mutual information, $I\left(\mathcal{S}:\mathcal{F}\right)/H(\mathcal{S})$.
For all figures, the parameters are $\Delta E=1$, $\delta\varepsilon=\Delta E$, $\lambda=\Delta E/5$, and $\beta=10$ for the thermal state ($\hbar=1$).  \label{fig:quantum_darwinism_perez_reduced}}
\end{figure*}

\begin{figure*}
\begin{centering}
\includegraphics[width=1\textwidth]{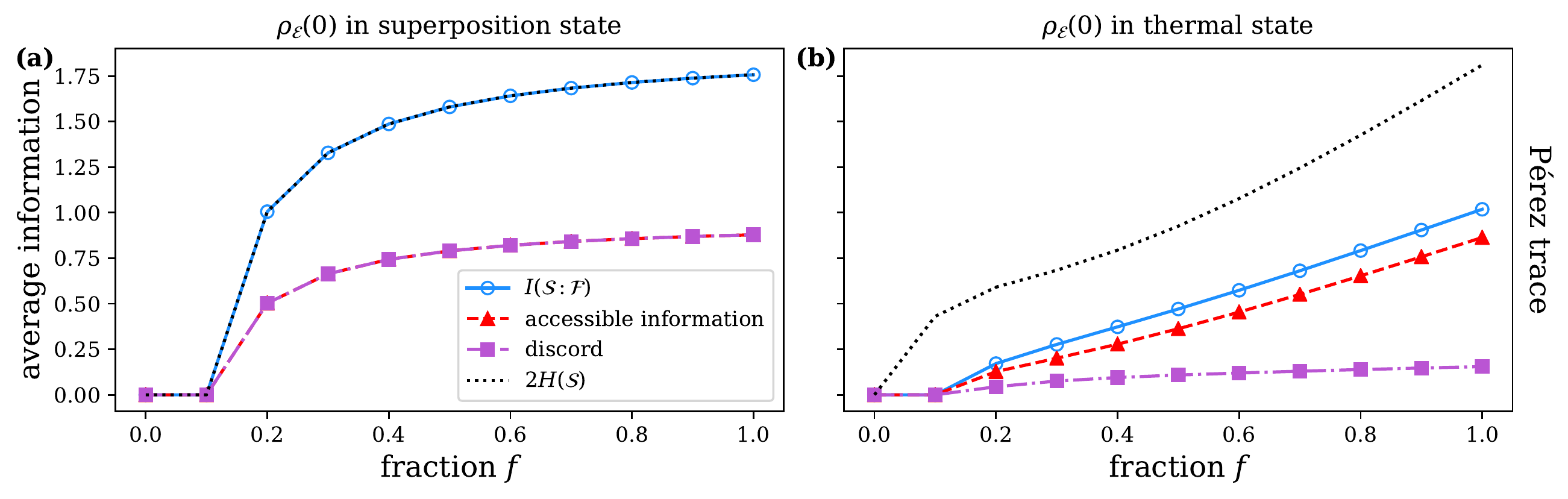}
\par\end{centering}
\caption{\textbf{Decomposition of mutual information $I(\mathcal{S}:\mathcal{F})$ into accessible information and quantum discord (at time $t=500$)  using the P\'erez partial trace  (Eq. (\ref{eq:PerezTrace}))} (compare with Fig.~\ref{fig:accessible_information}). The accessible information and quantum discord is averaged over all environment fractions of the same size $f$. The system entropy is $H(\mathcal{S})$, where the reduced system is calculated as $\rho_\mathcal{S}^\prime = \tr_\mathcal{F}\left[\rho_{\mathcal{SF}}\right]$. For all figures, the parameters are $\Delta E=1$, $\delta\varepsilon=\Delta E$, $\lambda=\Delta E/5$, and $\beta=10$ for the thermal state ($\hbar=1$). \label{fig:accessible_information_perez_reduced}}
\end{figure*}

The mutual information used in quantum Darwinism is given in Fig.~\ref{fig:quantum_darwinism_perez_reduced}, and the decomposition into accessible information and discord is given in Fig.~\ref{fig:accessible_information_perez_reduced} (the plot for the distance to spectrum broadcast structure does not employ the system entropy). Qualitatively, similar conclusions can be drawn from these results as was done in the main text. Notably for Fig.~\ref{fig:quantum_darwinism_perez_reduced}(a) and Fig.~\ref{fig:accessible_information_perez_reduced}(a), when the system starts off in a pure superposition state,  the P\'erez entropy of the system is \emph{equal} to the mutual information between system and fragment. This is due to the nature of the P\'erez trace, suggesting that the environment fragment always shares full information with the `false' system. Explicitly, $\rho_\mathcal{SF}$ are pure states under the P\'erez trace, and the reduced states $\rho_\mathcal{S}'$ and $\rho_\mathcal{F}$ are obtained from $\rho_\mathcal{SF}$  via the typical trace, hence the known property of pure states \cite{Blume-Kohout2005} applies, \emph{i.e.} that $I(\mathcal{S}:\mathcal{F})=2H(\mathcal{S})$, With reference to the discussion in Section \ref{subsec:qD_results}, the fact that the plots are non-symmetric follows.

\end{document}